\documentclass[useAMS,usenatbib]{mnras}
\usepackage{natbib}
\usepackage[english]{babel}
\usepackage{graphicx,amsmath,amsfonts,amssymb}
\usepackage{colortbl}
\usepackage{tabu,booktabs}
\usepackage{xcolor}
\usepackage{xspace}

\def \etal {et~al.~}

\newcommand{\hMpc}{{\ifmmode{h^{-1}{\rm Mpc}}\else{$h^{-1}$Mpc}\fi}}
\newcommand{\Mpc}{{\ifmmode{{\rm Mpc}}\else{Mpc}\fi}}
\newcommand{\hkpc}{{\ifmmode{h^{-1}{\rm kpc}}\else{$h^{-1}$kpc}\fi}}
\newcommand{\kpc}{{\ifmmode{ {\rm kpc} }\else{{\rm kpc}}\fi}}
\newcommand{\kms}{{\ifmmode{ {\rm km\,s^{-1}} }\else{ ${\rm km\,s^{-1}}$ }\fi}}
\newcommand{\hMsun}{{\ifmmode{h^{-1}{\rm {M_{\odot}}}}\else{$h^{-1}{\rm{M_{\odot}}}$}\fi}}
\newcommand{\Msun}{{\ifmmode{{\rm M}_{\odot}}\else{${\rm M}_{\odot}$}\fi}}
\newcommand{\Mhalo}{{\ifmmode{M_{\rm halo}}\else{$M_{\rm halo}$}\fi}}
\newcommand{\Rvir}{{\ifmmode{R_{\rm vir}}\else{$R_{\rm vir}$}\fi}}
\newcommand{\Mstar}{{\ifmmode{M_{\rm star}}\else{$M_{\rm star}$}\fi}}
\newcommand{\Vrot}{{\ifmmode{V_{\rm rot}}\else{$V_{\rm rot}$}\fi}}
\newcommand{\ltsima}{$\; \buildrel < \over \sim \;$}
\newcommand{\gtsima}{$\; \buildrel > \over \sim \;$}
\newcommand{\lsim}{\lower.5ex\hbox{\ltsima}}
\newcommand{\gsim}{\lower.5ex\hbox{\gtsima}}

\def\lesssim{\mathrel{\hbox{\rlap{\hbox{\lower4pt\hbox{$\sim$}}}\hbox{$<$}}}}
\def\gtrsim{\mathrel{\hbox{\rlap{\hbox{\lower4pt\hbox{$\sim$}}}\hbox{$>$}}}}

\newcommand{\beq}{\begin{equation}}
\newcommand{\eeq}{\end{equation}}
\def\beqa{\begin{eqnarray}}
\def\eeqa{\end{eqnarray}}
\def\LCDM{\ensuremath{\Lambda}CDM}

\def\head{ \vbox to 0pt{\vss \hbox to 0pt{\hskip 440pt\rm
      LA-UR-10-07069\hss} \vskip 25pt}}

\def \kms {\ifmmode  \,\rm km\,s^{-1} \else $\,\rm km\,s^{-1}  $ \fi }
\def \kpc {\ifmmode  {\rm kpc}  \else ${\rm  kpc}$ \fi  }  
\def \hkpc {\ifmmode  {h^{-1}\rm kpc}  \else ${h^{-1}\rm kpc}$ \fi  }  
\def \hMpc {\ifmmode  {h^{-1}\rm Mpc}  \else ${h^{-1}\rm Mpc}$ \fi  }  
\def \Mpch {\ifmmode  {h^{-1}\rm Mpc}  \else ${h^{-1}\rm Mpc}$ \fi  }  
\def \Msun {\ifmmode {\rm M}_{\odot} \else ${\rm M}_{\odot}$ \fi} 
\def \hMsun {\ifmmode h^{-1}\,\rm M_{\odot} \else $h^{-1}\,\rm M_{\odot}$ \fi}

\def \LCDM {\ifmmode \Lambda{\rm CDM} \else $\Lambda{\rm CDM}$ \fi}
\def \sig8 {\ifmmode \sigma_8 \else $\sigma_8$ \fi} 
\def \OmegaM {\ifmmode \Omega_{\rm m} \else $\Omega_{\rm m}$ \fi} 
\def \Omegab {\ifmmode \Omega_{\rm b} \else $\Omega_{\rm b}$ \fi} 
\def \OmegaL {\ifmmode \Omega_{\rm \Lambda} \else $\Omega_{\rm \Lambda}$\fi} 
\def \Deltavir {\ifmmode \Delta_{\rm vir} \else $\Delta_{\rm vir}$ \fi}
\def \rhocrit {\ifmmode \rho_{\rm crit} \else $\rho_{\rm crit}$ \fi}
\def \rhou {\ifmmode \rho_{\rm u} \else $\rho_{\rm u}$ \fi}
\def \zc {\ifmmode z_{\rm c} \else $z_{\rm c}$ \fi}

\def\lcdm{\ensuremath{\Lambda\textrm{CDM}}\xspace}

\def\aj{AJ}%
\def\araa{ARA\&A}%
\def\apj{ApJ}%
\def\apjl{ApJ}%
\def\apjs{ApJS}%
\def\aap{A\&A}%
\def\mnras{MNRAS}%
\def\nat{Nature}%
\def\head{ .ps \vbox to 0pt{\vss \hbox to 0pt{\hskip 440pt\rm
      LA-UR-10-07069\hss} \vskip 25pt}} 

\def \spose#1{\hbox  to 0pt{#1\hss}}  
\def \lta{\mathrel{\spose{\lower 3pt\hbox{$\sim$}}\raise 2.0pt\hbox{$<$}}}
\def \gta{\mathrel{\spose{\lower 3pt\hbox{$\sim$}}\raise 2.0pt\hbox{$>$}}}

\title[Constraining star formation parameters]
{An observational test for star formation prescriptions in cosmological hydrodynamical simulations}

\author[Buck \etal] {Tobias Buck$^{1,2}$\thanks{E-mail:
    buck@mpia.de}, 
    Aaron A. Dutton$^{3}$,
    Andrea V. Macci\`o$^{3,2}$ \\
$^1$Leibniz-Institut f\"ur Astrophysik Potsdam (AIP), An der Sternwarte 16, D-14482 Potsdam, Germany\\
$^2$Max-Planck-Institut f\"ur Astronomie, K\"onigstuhl 17, 69117 Heidelberg, Germany\\
$^3$New York University Abu Dhabi, PO Box 129188, Saadiyat Island, Abu Dhabi, United Arab Emirates
}

\begin{document}

\date{Accepted 2019 April 02. Received 2019 March 31; in original form 2018 December 13}

\pagerange{\pageref{firstpage}--\pageref{lastpage}} \pubyear{2019}

\maketitle

\label{firstpage}


\begin{abstract}
State-of-the-art cosmological hydrodynamical simulations of galaxy formation have reached the point at which their outcomes result in galaxies with ever more realism. Still, the employed sub-grid models include several free parameters such as the density threshold, $n$, to localize the star-forming gas. In this work, we investigate the possibilities to utilize the observed clustered nature of star formation (SF) in order to refine SF prescriptions and constrain the density threshold parameter. To this end, we measure the clustering strength, correlation length and power-law index of the two-point correlation function of young ($\tau<50$ Myr) stellar particles and compare our results to observations from the HST Legacy Extragalactic UV Survey (LEGUS). Our simulations reveal a clear trend of larger clustering signal and power-law index and lower correlation length as the SF threshold increases with only mild dependence on galaxy properties such as stellar mass or specific star formation rate. In conclusion, we find that the observed clustering of SF is inconsistent with a low threshold for SF ($n<1$ cm$^{-3}$) and strongly favours a high value for the density threshold of SF ($n>10$ cm$^{-3}$), as for example employed in the NIHAO project.
\end{abstract}

\noindent
\begin{keywords}

  cosmology: dark matter - galaxies: formation - galaxies: structure - galaxies: star formation - ISM: structure - methods: N-body simulation

 \end{keywords}

\vspace*{-1.5cm}
\section{Introduction} \label{sec:intro}

It is widely assumed that stars primarily form in clusters across the stellar disks of galaxies. Because current cosmological simulations barely resolve the sites and processes of star formation (SF) itself, most state-of-the-art simulations implement phenomenological recipes for SF on scales of giant molecular clouds (GMCs).
Although little is known about the small scale details of the SF process, such as the collapse and onset of GMC fragmentation or the impact of turbulent and magnetic pressure on the gas clouds, the larger scale galactic effects are well studied and understood \citep{Kennicutt2012}. E.g. the typical masses of GMCs, the sites of SF, are $\sim10^2 - 10^4\Msun$ with typical densities of $\sim10^2-10^5$ cm$^{-3}$. 

In order to identify analogues of molecular clouds in state-of-the-art hydrodynamical simulations several recipes have been introduced \citep[see e.g.][] {Hopkins2013}. Due to the huge dynamical ranges involved in the formation of galaxies (from kilo-parsec scales down to sub-parsec scales) several key physical ingredients need to be modelled as sub-grid models with free parameters. One of these sub-grid models deals with the formation of stars out of the interstellar gas where SF is usually localized via a density threshold $n$. This threshold varies over more than 5 orders of magnitude across different simulations from $n=0.01-10^3$ cm$^{-3}$. 
Only very few works focused on the impact of different localization criteria for SF on the global galaxy properties \citep[][]{Hopkins2013,Dutton2018}  and even fewer works compared the detailed outcome of the adopted SF criteria to the observed properties of star forming regions in galaxies. 
 
Although current cosmological simulations lack the resolution to resolve the details of SF inside the molecular clouds, the large scale effects and properties of these clouds are well resolved in the highest resolution simulations today, e.g. APOSTLE \citep{Sawala2016}, AURIGA \citep{Grand2017}, FIRE \citep{Hopkins2018} or NIHAO \citep{Buck2018c} and especially the observed spatial clustering behaviour of young stellar complexes \citep{Grasha2015,Grasha2017} should be well resembled by current models of SF.

\begin{figure*}
\begin{center}
\includegraphics[width=\textwidth]{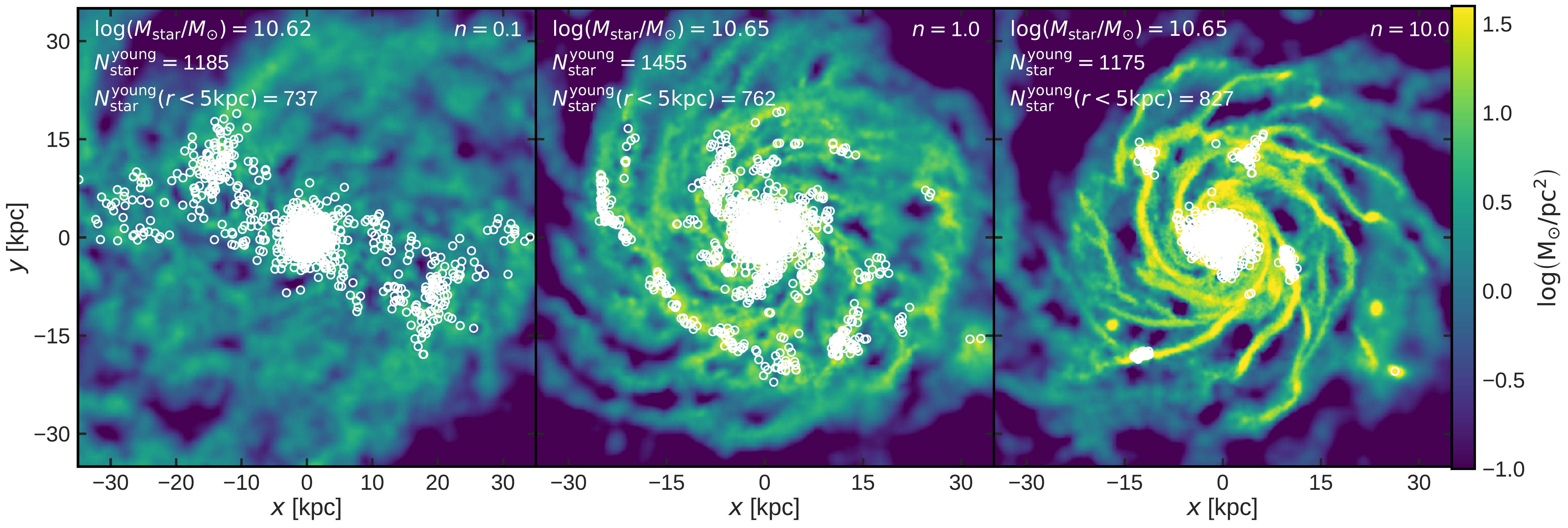}
\end{center}
\vspace{-.35cm}
\caption{Spatial distribution of young star particles for the three different threshold runs of the galaxy g8.26e11 at redshift $z=0.13$. Colour-coding shows the surface density of neutral hydrogen gas and open white circles the position of star particles younger than 50 Myr. Note the on average much higher gas densities achieved in progressively higher threshold runs.}
\label{fig:young_stars}
\end{figure*}

Here we set out to introduce and describe a conclusive test for SF models in cosmological simulations by means of the clustering signal of young star clusters (in both simulations and observations). We use the results on clustered SF to put constraints on (up to now) free parameters of current cosmological simulations of galaxy formation. This paper is structured as follows: in \S2 we describe the simulation suite, in \S3 we lay out the methodology used in this work, in \S4 we present our constraints on the SF parameter and finally in \S5 we discuss our results and present our conclusions.

\section{Cosmological Simulations} \label{sec:sim}

For this work we use 17 sets of simulations with stellar masses from $10^7 \lesssim M_{\rm star} \lesssim10^{11}$ \Msun. For each simulation we analyze the most massive galaxy at 13 snapshots equally spaced in time between redshift $z=0$ and $z=0.5$. 

These simulations are a sub-sample of the NIHAO hydrodynamical cosmological zoom-in suite \citep{Wang2015} run with the smooth-particle-hydrodynamics (SPH) code \texttt{Gasoline2} \citep{Wadsley2017}. The galaxy formation model is described in detail in the papers of \citet{Wang2015} and \citet{Dutton2018}. The most important parameter to our study is the SF threshold, $n$, which we describe below. 

The fiducial NIHAO simulations adopt as SF threshold of $n = 10$ cm$^{-3} \sim 50 m_{\rm gas}/\epsilon_{\rm gas}^3$ where 50 denotes the number of neighbours in the SPH smoothing kernel, $m_{\rm gas}$ the initial gas particle mass, and $\epsilon_{\rm gas}$ the gravitational force softening \citet{Stinson2006,Stinson2013}. In NIHAO $\epsilon_{\rm gas} \propto m_{\rm gas}^{1/3}$, so that $n$ is independent of gas particle mass. Each fiducial simulation is complemented by three additional SF threshold runs: $n = 0.1$, $n = 1.0$ and $n=20.0$. The former two are further described and used in \citet{Dutton2018}. The threshold of $n = 0.1$ is similar to the values adopted by the EAGLE/APOSTLE \citep{Schaye2015,Sawala2016} and ILLUSTRIS/AURIGA \citep{Vogelsberger2014,Grand2017} projects\footnote{Note: All values of $n$ used here are well below typical densities of GMCs ($n\sim100$ cm$^{-3}$) and their SF cores ($n \sim 10^4$ cm$^{-3}$).}. While the SF threshold is not the only parameter than can be varied, in this study we focus on this particular parameter given its strong impact on galaxy morphology and halo response \citep[e.g.][]{Dutton2018,Benitez2018,Bose2018}.
We like to highlight that the $n=20.0$, $n=10.0$ and $n=1.0$ simulations are run with exactly the same parameters except for the star formation threshold \citep[see][for more details]{Dutton2018} and still follow the same abundance matching relation across cosmic time although with very different morphology for the young stars. The $n=0.1$ runs were recalibrated by reducing the efficiency of the early stellar feedback from $e=0.13$ to $0.04$ in order to make the galaxies fit the abundance matching relation again.

Haloes in the zoom-in regions are identified using the halo finder \texttt{AHF2} \citep{Knollmann2009,Gill2004}. The virial mass, $M_{200}$, denotes the mass of all particles within a sphere of virial radius, $R_{200}$, containing $\Delta$ = 200 times the cosmic critical matter density, $\rhocrit$. In general, galaxy properties are measured from all hydro particles within $0.2 R_{200}$. Dark matter particle masses and force softenings are chosen to resolve the mass profile at $\lesssim1\%$ of $R_{200}$ with dark matter force softenings of $\epsilon_{\rm dark} = 207-931$ pc, and hydro softenings of $\epsilon_{\rm gas} = 88-400$ pc. Gas particle masses range between $3.5\times10^3\Msun-3.2\times10^6\Msun$\footnote{The results presented in this work have been compared against the simulations with 8 times higher mass resolution presented in \citet{Buck2018c} and we found good convergence of the clustering behaviour of young stars.}. All simulations in the NIHAO project, including the ones used here, employ a pressure floor to keep the Jeans mass of the gas resolved and suppress artificial fragmentation \citep[see also appendix A1 of][]{Smith2018}. Our implementation follows \citet{Agertz2009} which is equivalent to the criteria proposed in \citet{Richings2016} and fulfils the \citet{Truelove1997} criterion at all times. Thus, the Jeans mass in our simulations is resolved with $\sim 4$ SPH kernel masses.

\begin{figure*}
\begin{center}
\includegraphics[width=\textwidth]{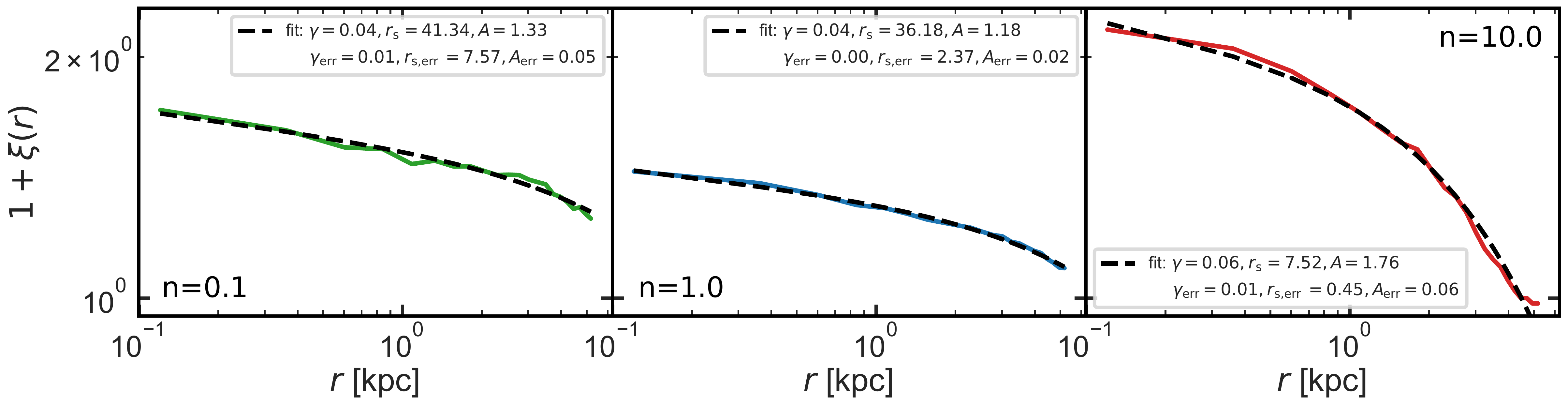}
\end{center}
\vspace{-.35cm}
\caption{Correlation function, $\xi(r)$, of star particles younger than 50 Myr in the simulation g3.55e11 for the three different threshold runs at redshift $z\sim0.1$. The solid colored lines show the correlation function measured from the simulation with the threshold value indicated in the panels. The dashed black line shows the truncated power law fit with the fit parameters given in the legend.}
\label{fig:ex}
\end{figure*}

NIHAO galaxies provide a perfect test sample to study the effects of the SF parameters on the structure of galaxies since \textit{all simulations} used here (runs for $n=0.1, 1.0, 10.0$) follow the abundance matching relations since $z = 4$ \citep[see][]{Dutton2018}. They form the right amount of stars \citep{Wang2015} and they follow the cold gas versus stellar mass relation at $z = 0$ \citep{Stinson2015,Buck2017}. Furthermore, they are in very good agreement with the Local Group stellar mass function \citep{Buck2018c} as well as the velocity function \citep{Maccio2016,Dutton2018a} and form MW analogues matching the the detailed kinematics of the MW bulge \citep{Buck2018a}.

\section{Method: Two-Point Correlation function of young star clusters} \label{sec:method}

We use the two-point correlation function (TPCF), $\xi(r)$, to measure the magnitude of clustering of young stellar particles as a function of separation $r$ in a suite of cosmological hydrodynamical simulations. The two-dimensional correlation function $1 + \xi(r)$ is defined as the probability above Poisson noise of finding two stellar particles with a separation $r$ as $dP = N^2[1 + \xi(r)] d\mathbf{x_1}d\mathbf{x_2}$, where N is the surface density of young stellar particles within two infinitesimal spatial elements $d\mathbf{x_1}$ and $d\mathbf{x_2}$, separated by distance $r$ \citep[see e.g.][for a similar definition in spherical coordinates]{Peebles1980}. For a truly random Poisson distribution, the TPCF will be flat across all separations, such that $1+\xi(r) = 1$. On the other hand, a clustered stellar distribution will have $1 + \xi(r) > 1$ at small values of $r$ and will be declining with increasing $r$ towards that of a flat, non-clustered distribution. The correlation function of a fractal (self-similar) distribution is described with a single power-law as $1 + \xi(r) = (r/r_{\rm s})^{-\gamma}$, where $r_{\rm s}$ is the correlation length of the clustering and $\gamma$ describes the hierarchical ordering \citep{Calzetti1989,Larson1995}.

In a two-dimensional self-similar distribution, the total number of clusters $N$ increases with radius $r$ as $N \propto r^{D_2}$, where $D_2$ is the two-dimensional fractal dimension \citep{Mandelbrot1982}. The number of clusters for every radial aperture will increase as $N \propto r^{-\gamma} \times r^2 \propto r^{-\gamma+2}$. Thus, the power-law slope $\gamma$ of the correlation function $1 + \xi(r)$ determines the two-dimensional fractal dimension as $D_2 = -\gamma + 2$. A flat, non-clustered distribution of $\gamma = 0$ will result in a fractal geometric dimension of $D_2 = 2$; a steep slope will indicate a clustered distribution with a fractal dimension less than 2. Indeed, the distribution of SF and interstellar gas is shown to exhibit a projected, two-dimensional fractal dimension of $D_2 \sim 1.2-1.6$ over a large range of environments \citep[e.g.][]{Sanchez2005,Elmegreen2006,Sanchez2008}. 

We implement the calculation of the TPCF using the standard estimator formula:
\begin{equation}\label{eq:LS}
\xi(r) = \frac{DD(r)}{RR(r)} -1
\end{equation}
Here $DD$ denotes the number of data-data pairs and $RR$ the number of random-random pairs with separation between $r$ and $r + \delta r$.
From this we determine the slope and the correlation length of the TPCF by fitting a power-law with an exponential cut-off to account for the finite extension of the stellar disc in the form of:
\begin{equation}
    1+\xi(r)=A\left(\frac{r}{r_{\rm s}}\right)^{-\gamma} \exp\left(-\frac{r}{r_{\rm s}}\right)
    \label{eq:power}
\end{equation}
$A$ is a normalization parameter, $r_{\rm s}$ the scale radius and $\gamma$ the power law index.

Recent measurements of the clustering of young stellar complexes are obtained from the HST Legacy Extragalactic UV Survey \citep[][LEGUS]{Calzetti2015,Adamo2017,Messa2018}. \citet{Grasha2017} measure the TPCF of young stellar complexes at scales from $5$ pc to $10$ kpc and fit it with a double power law measuring the correlation length, clustering strength and the power law index\footnote{Note: We define $\gamma$ without the minus sign which leads to a sign flip between our results and the ones of \citet{Grasha2017}.} below and above the break radius of $r\sim100$ pc.

So in what follows we use their average power-law fit (from their table 4) at large radii in the form of $1+\xi(r) = 5.6(\pm0.3) r^{-0.21\pm0.02}$ to compare with NIHAO. We use the results from \citet{Grasha2017} for young star clusters with ages of $\lesssim40$ Myr and sizes of roughly $\lesssim300$ pc (their classes 1, 2, 3) which is well in agreement with the spatial extent of stellar tracer particles in our cosmological simulations.
From the fits of \citet{Grasha2017} and the fits to our simulation data we derive the clustering strength at a scale of $r=100$ pc, $1 + \xi(r=100\rm{pc})$, and the correlation length as the radius where the correlation function $\xi$ goes to zero, $r(\xi=0)$ and compare the two results. A complete list of fitted and derived clustering parameters for the simulations and observations is given in table \ref{tab:fits}. 

\begin{table}
\caption{Clustering parameters for young stellar particles: For each threshold run we show from top to bottom the three fit parameters, scale radius, $r_{\rm s}$, power-law index $\gamma$ and normalization, $A$, as well as derived values for the radius $r(\xi=0)$ where the clustering strength goes to zero and the value of the clustering strength measured at a radius of $r=100$pc, $1 + \xi(r=100\rm{pc})$. From left to right we show the median with uncertainty ranges, 16$^{\rm th}$ and 84$^{\rm th}$ percentile and the sample size. Note: The sample size varies because some snapshots are discarded due to too few young stellar particles.}
\label{tab:fits}
\begin{tabular}{l c c c c}
		\hline\hline
		parameter & median & $p^{16th}$ & $p^{84th}$ & $N_{\rm gal}$ \\
		\hline
	 \multicolumn{5}{c}{$n=20.0$}\\
        \hline
		$r_{\rm s}$ [kpc] & $3.55^{+0.71}_{-0.17}$ & 1.67 & 11.62 & 189\\
		$\gamma$ & $0.074^{+0.010}_{-0.011}$ & -0.051 & 0.192 & 189\\
		$A$ & $ 1.57^{+0.15}_{-0.06}$ & 0.89 & 3.24 & 189\\
		$r(\xi=0)$ [kpc] & $1.38^{+0.10}_{-0.12}$ & 1.13 & 2.53 & 189\\
		$1 + \xi(r=100\rm{pc})$ & $ 2.19^{+0.13}_{-0.07}$ & 1.32 & 3.68 &  189\\
        \hline
        \multicolumn{5}{c}{$n=10.0$}\\
        \hline
		$r_{\rm s}$ [kpc] & $4.77^{+0.54}_{-0.26}$ & 2.13 & 10.13 & 148\\
		$\gamma$ & $0.094^{+0.012}_{-0.007}$ & 0.028 & 0.213 & 148\\
		$A$ & $1.31^{+0.07}_{-0.03}$ & 0.97 & 1.96 & 148\\
		$r(\xi=0)$ [kpc] & $1.84^{+0.10}_{-0.09}$ & 0.97 & 2.87 & 148\\
		$1 + \xi(r=100\rm{pc})$ & $2.07^{+0.13}_{-0.05}$ & 1.53 & 3.38 & 148 \\
        \hline
        \multicolumn{5}{c}{$n=1.0$}\\
        \hline
		$r_{\rm s}$ [kpc] & $15.45^{+2.24}_{-0.81}$ & 6.84 & 39.25 & 168\\
		$\gamma$ & $0.032^{+0.007}_{-0.003}$ & 0.003 & 0.106 & 168 \\
		$A$ & $1.17^{+0.06}_{-0.02}$ & 0.92 & 1.82 & 168\\
		$r(\xi=0)$ [kpc] & $2.69^{+0.35}_{-0.19}$ & 0.67 & 6.36 & 168\\
		$1 + \xi(r=100\rm{pc})$ & $1.45^{+0.07}_{-0.02}$ & 1.23 & 2.15 & 168\\
        \hline
        \multicolumn{5}{c}{$n=0.1$}\\
        \hline
		$r_{\rm s}$ [kpc] & $10.79^{+0.85}_{-0.34}$ & 7.05 & 20.17 & 180\\
		$\gamma$ & $0.027^{+0.004}_{-0.003}$ & -0.009 & 0.076 & 180\\
		$A$ & $1.35^{+0.06}_{-0.02}$ & 1.10 & 1.98 & 180\\
		$r(\xi=0)$ [kpc] & $3.07^{+0.21}_{-0.16}$ & 1.27 & 5.44 & 180\\
		$1 + \xi(r=100\rm{pc})$ & $1.63^{+0.05}_{-0.03}$ & 1.32 & 2.16 & 180 \\
        \hline
        \multicolumn{5}{c}{observations}\\
        \hline
		$\gamma$ &  \multicolumn{3}{c}{$0.21\pm0.02$} &\\
		$r(\xi=0)$ [kpc] &  \multicolumn{3}{c}{$1.44\pm0.83$} & \\
		$1 + \xi(r=100\rm{pc})$ &  \multicolumn{3}{c}{$2.13\pm0.23$} & \\
        \hline
\end{tabular}
\end{table}

\begin{figure*}
\begin{center}
\includegraphics[width=\textwidth]{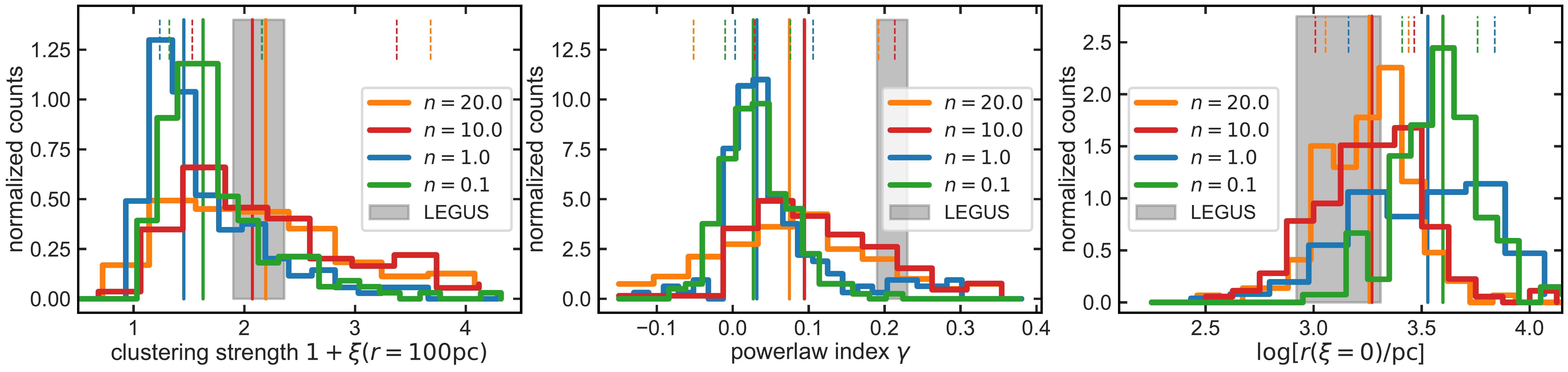}
\end{center}
\vspace{-.35cm}
\caption{Distribution of the clustering strength $1 + \xi(r=100\rm{pc})$ (left panel), power law index $\gamma$ (middle panel) and correlation length $r(\xi=0)$ for the set of simulations. The low threshold ($n=0.1$) run is shown in green, the $n=1.0$ run in blue, the fiducial $n=10.0$ run with red and the $n=20.0$ run in orange. Vertical solid and dashed lines show the median and the $16^{\rm th}$ and $84^{\rm th}$ percentile ranges respectively. Observational constraints from the LEGUS survey \citep{Grasha2017} are shown in gray bands.}
\label{fig:dist}
\end{figure*}

\section{Results: Constraints on the star formation threshold}
\label{sec:results}

\begin{figure*}
\begin{center}
\includegraphics[width=\textwidth]{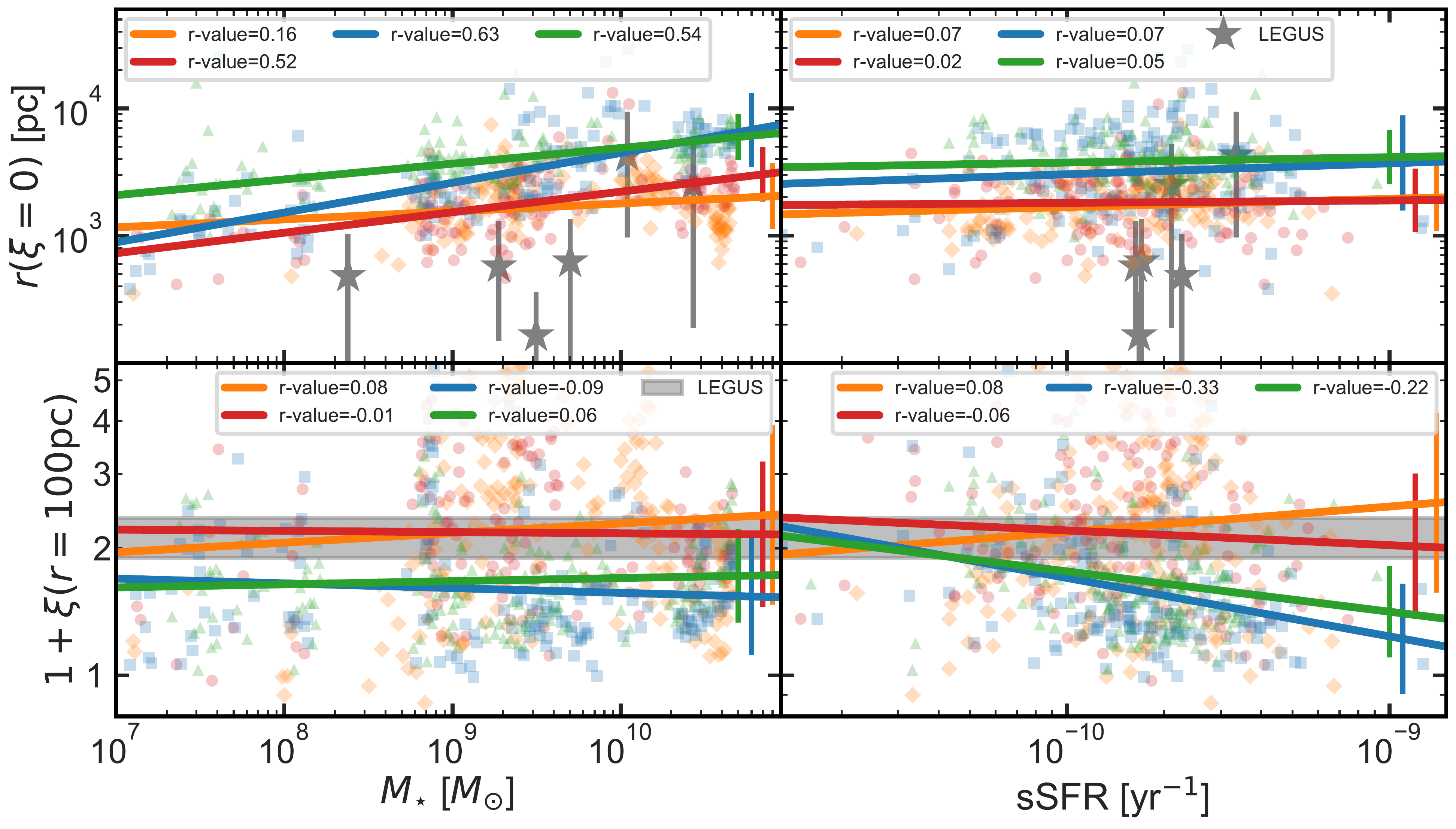}
\end{center}
\vspace{-.35cm}
\caption{Correlation length $r(\xi=0)$ and clustering strength $1 + \xi(r=100\rm{pc})$ as a function of stellar mass and specific SFR. The low threshold  run ($n=0.1$) is shown with green triangles, the $n=1.0$ run with blue squares, the fiducial high threshold run ($n=10.0$) with red dots and the $n=20$ runs with orange diamonds. Observational results from the LEGUS survey are shown with either gray stars where values for single galaxies can be found or gray bands (where observations only give averaged values) respectively. For each threshold run we test for a linear correlation reporting the correlation coefficient in the legend and indicating the $1\sigma$ scatter around the mean relation with thin vertical lines}.
\label{fig:4_panel}
\end{figure*}

In figure \ref{fig:young_stars} we show three versions of the same galaxy simulated with different density threshold values increasing from left to right. Open white circles show the positions of young stars ($\tau<50$ Myr) and color-coding shows the surface density of neutral hydrogen gas computed following the results from \citet{Rahmati2013} as implemented by \citet{Gutcke2017}. Their total number as well as their number in the inner 5 kpc is indicated in each panel and is roughly equal for all three runs. We see that the high threshold run forms stars that are more clustered compared to the lower threshold runs. Especially the equal number of young stars within the central region shows that this not due simply more SF in the very center of the galaxy. This figure confirms the naive expectation that a higher threshold might lead to a more clustered SF compared to a low threshold. In the following section we analyse the effects of the SF threshold on the spatial distribution of young stars by fitting equation \ref{eq:power} to the TPCF of young stellar particles derived from the simulations.
An example of this power law fit to the TPCF derived from one of our galaxies is shown in figure \ref{fig:ex} with the fit parameters and their uncertainties shown in the legend. Typical fitting uncertainties on the derived parameters ($r_{\rm s}$ and $\gamma$) are on the order of $\lesssim20\%$ for $\gamma$ and $\lesssim5\%$ for $r_{\rm s}$. A stronger correlation of young stars in the high threshold run can be appreciated from the shorter correlation length, $r_{\rm s}$, and the overall stronger correlation signal compared to the other two threshold runs.

In figure \ref{fig:dist} we compare the distribution of the clustering strength $1 + \xi(r=100\rm{pc})$, the power-law index $\gamma$ and the correlation length $r(\xi=0)$ for our four sets of simulations. The $n=0.1$ runs are shown with green histograms, the $n=1.0$ runs in blue, the $n=10.0$ ones with red histograms and the $n=20.0$ runs with orange histograms. Vertical coloured lines indicate the median values and gray vertical bands show the observations. There is a clear trend of larger clustering strength and power-law index and lower correlation length with increasing SF threshold. While for all simulations the power-law index $\gamma$ is lower than observed, the clustering strength and the correlation length of the high ($n=10$) SF threshold runs are in agreement with the observed results. This shows that indeed SF is more clustered in the high threshold simulations.

\subsection{Clustered star formation}
 
We show the results for the correlation length, $r(\xi=0)$, and the clustering strength, $1 + \xi(r=100\rm{pc})$, derived from the fits of \citet{Grasha2017} for a set of 6 LEGUS galaxies with gray stars or gray shaded bands in the figures.

In figure \ref{fig:4_panel} we compare the correlation length (upper panels) and the clustering strength (lower panels) as a function of stellar mass (left panels) and specific SFR (sSFR, right panels) to observational results. We show with gray stars the results for individual galaxies where present (upper panels) and with gray bands the averaged observational results for the six LEGUS galaxies as presented in \citet[][lower panels]{Grasha2017}. For our simulations we further show a linear regression to test for any correlation and display the correlation coefficients in the legend.  
In general, we recover the findings from figure \ref{fig:dist} that the two lower threshold runs exhibit both larger correlation length and smaller clustering strength. We find a weak correlation of $r(\xi=0)$ with stellar mass and a weak anti-correlation of $1 + \xi(r=100\rm{pc})$ with sSFR.
We find that $r(\xi=0)$ derived from the simulations is slightly larger compared to the observed values of $\sim1$ kpc but consistent with the observed values (gray stars) for the largest stellar masses and the highest threshold runs (red). On the other hand, for the $n=10.0$ runs $1 + \xi(r=100\rm{pc})$ is in agreement with the observed range of $\sim1.90-2.36$ (gray bands). The $n=0.1$ (green) and $n=1.0$ (blue) runs are inconsistent with the observed values for both $r(\xi=0)$ and $1 + \xi(r=100\rm{pc})$ although we note there is a large scatter.

\begin{figure}
\begin{center}
\includegraphics[width=\columnwidth]{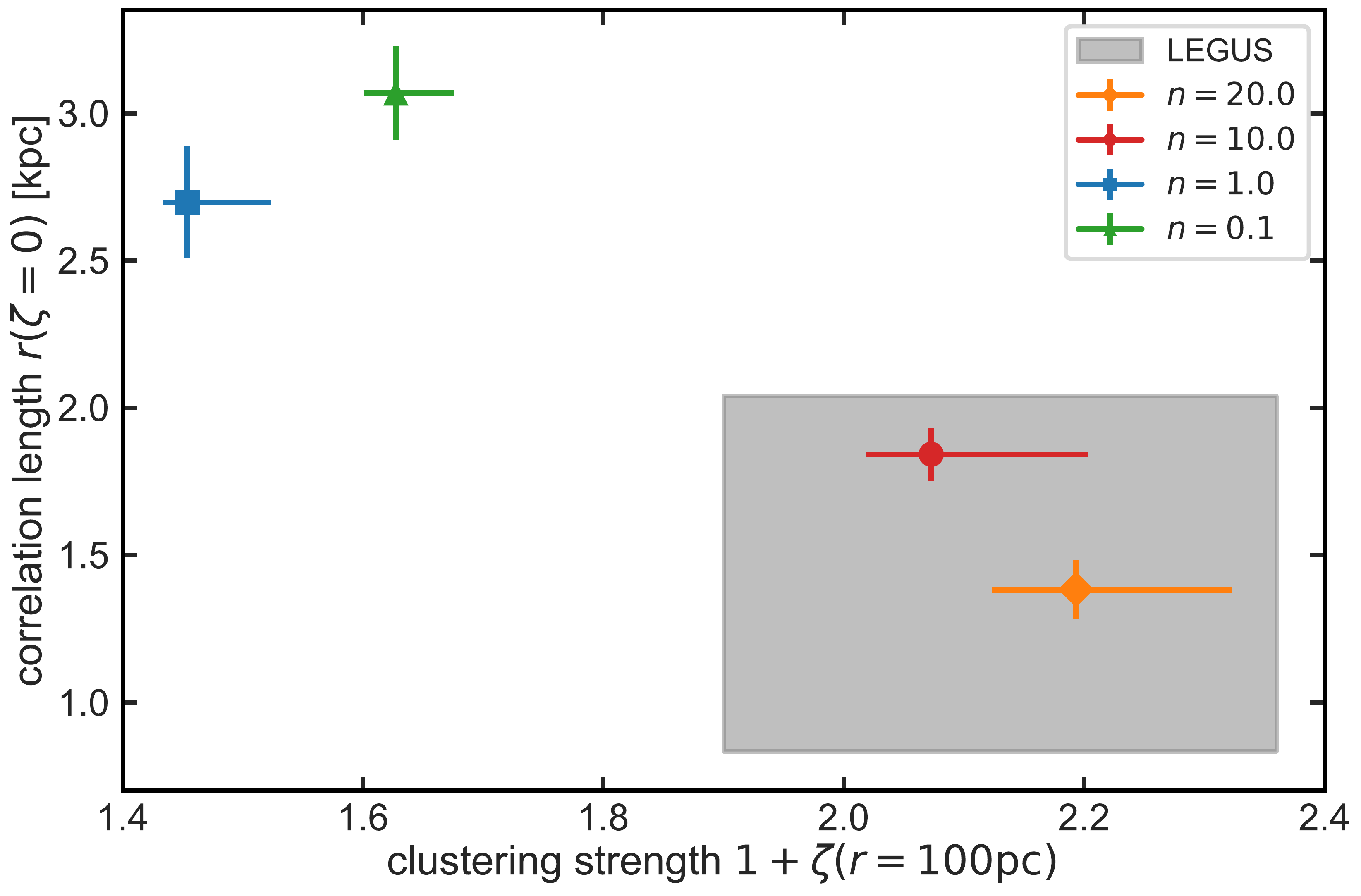}
\end{center}
\vspace{-.35cm}
\caption{Correlation length $r(\xi=0)$ vs clustering strength $1 + \xi(r=100\rm{pc})$. Coloured points show the simulation median with error bars indicating the uncertainty on the median ($\sqrt{N_{\rm gal}}\sim10$ times smaller than the scatter). The gray box shows the observed parameter range from \citet{Grasha2017}.}
\label{fig:gamma_r0}
\end{figure}

These findings are further supported by figure \ref{fig:gamma_r0} which plots the correlation length $r(\xi=0)$ vs. the clustering strength, $1 + \xi(r=100\rm{pc})$. For the simulations we show median values over all galaxies and redshifts. The error bars show the uncertainty on the median defined as: $\sigma_{\rm high}=1.22 (p_{84}-p_{50})/\sqrt {N_{\rm gal}}$ and  $\sigma_{\rm low}=1.22 (p_{50}-p_{16})/\sqrt {N_{\rm gal}}$ respectively. Here $p_{84}$, $p_{16}$ and $p_{50}$ denote the $84^{\rm th}$, $16^{\rm th}$ percentile and the median. $N_{\rm gal}$ denotes the number of galaxies. The observed parameter range is shown with a gray box. The $n=0.1$ and $n=1.0$ runs are located in the upper left corner of this plot while the $n=10.0$ and $n=20.0$ runs show smaller correlation length and larger clustering strength, respectively. This indicates that only the high threshold runs ($n=10.0$ and $n=20.0$) result in a clustering of young stellar particles which is consistent with the observed clustering of young star clusters. This diagram shows a trend of increasing clustering of SF with increasing density threshold $n$ and thus advocates for the usage of a high density threshold in simulations of galaxy formation; much higher than what is used in the AURIGA or APOSTLE simulations.

\section{Discussion and Conclusion} \label{sec:dis}


State-of-the-art cosmological hydrodynamical simulations of galaxy formation have reached the point at which their outcomes result in galaxies with ever more realistic properties \citep[e.g.][]{Buck2018c,Grand2017}. Still, the employed models are rather simplistic and involve several free parameters. One of these parameters is the density threshold $n$ used to localize the SF gas in the simulation. This parameter varies over several orders of magnitude among different simulation groups.

In this work we set out to investigate the prospects of using the observed clustered nature of SF in order to refine the SF prescriptions in simulations and constrain the density threshold parameter. We find that the observed strong clustering of SF only agrees with simulations that employ a high threshold for star formation ($n\gsim10$) like e.g. the NIHAO \citep{Wang2015} and FIRE \citep{Hopkins2018} simulations, the super-bubble feedback runs of \citet{Keller2014} or simulations using an H$_2$ based SF prescription which naturally leads to a high density for star forming gas \citep[e.g.][]{Christensen2012,Munshi2018}. Such high SF thresholds runs have been used to simulate realistic bulgeless galaxies \citep{Governato2010} and dwarf galaxy populations of Milky Way mass galaxies \citep{Zolotov2012,Brooks2014}. 
On the other hand, observations disfavour simulations with a low density threshold like the AURIGA/ILLUSTRIS \citep{Grand2017,Vogelsberger2014,Pillepich2018}, APOSTLE/EAGLE \citep{Sawala2016,Schaye2015}, ROMULUS \citep{Tremmel2017} or VELA \citep{Zolotov2015} simulations. A lower value of the star formation threshold, $n$, results in a significant decrease in simulation runtime but comes at the cost of a less realistic star formation model. Furthermore, $n$ strongly impacts the predicted dark matter halo structure. As recently shown by \citet{Dutton2018} a high value of $n$ naturally resolves the Too-Big-Too-Fail and the cusp-core tension of \lcdm. On the other hand, this implies that a low value of $n$ does not lead to accurate predictions for the structure of \lcdm haloes. 

In this study we focussed on varying the star formation threshold keeping other parameters like e.g. the star formation efficiency and the number of SPH kernel masses in the pressure floor fixed. Thus, it remains to be seen whether the clustering of the young stars for low star formation thresholds can be increased by choosing different numerical values for these parameters. However, blindly varying parameters does not teach us something about the underlying physics of galaxy formation. We thus see the presented method as an observational test of future star formation prescriptions in cosmological simulations. Such models have to recover not only the star formation histories and the total stellar masses (integrated star formation histories) but also the observed spatial distribution of star formation itself.

Finally, while it is a priori evident that a simple threshold value must be a too simplistic description of the SF process, our current computational resources prohibit more complex first principle prescriptions, at least for fully cosmological simulations. Improved models for isolated galaxy simulations without cosmological context exist \citep[][]{Hu2017,Semenov2017,Emerick2018} and updated models for cosmological simulations are explored \citep{Applebaum2018}. Therefore, we believe that the presented correlations between spatial clustering of young stellar complexes and the density threshold can be used to gauge the value of $n$ in future simulations and remove one free parameter from the models. Furthermore, given the current successes in galaxy formation models and the wealth of observational data obtained from large scale surveys the onus is now to compare model predictions and observations beyond simple integral properties such as total stellar mass or galaxy size and use the full range of galactic morphologies to test numerical and theoretical models. Only in this way we will improve our understanding of galaxy formation.

\section*{Acknowledgments}
We thank the referee for an insightful report which helped us improve the manuscript. 
TB acknowledges support from the Sonderforschungsbereich SFB 881 “The Milky Way System” (subproject A2) of the German Research Foundation (DFG) and by the European Research Council under ERC-CoG grant CRAGSMAN-646955. The authors gratefully acknowledge the Gauss Centre for Supercomputing e.V. (www.gauss-centre.eu) for funding this project by providing computing time on the GCS Supercomputer SuperMUC at Leibniz Supercomputing Centre (www.lrz.de). This research was carried out on the High Performance Computing resources at New York University Abu Dhabi; Simulations have been performed on the ISAAC cluster of the Max-Planck-Institut f\"ur Astronomie and the HYDRA and DRACO clusters at the Rechenzentrum in Garching. This research made use of the {\sc{pynbody}} \citet{pynbody}, {\sc{tangos}} \citet{tangos}, {\sc{matplotlib}} \citep{matplotlib}, {\sc{SciPy}} \citep{scipy} and {\sc{NumPy, IPython and Jupyter}} \citep{numpy,ipython,jupyter} {\sc{python}} packages. 



\bibliography{astro-ph.bib}

\label{lastpage}

\end{document}